\documentclass[a4paper,manyauthors,nocleardouble,COMPASS]{cernphprep}

\usepackage{bm}
\usepackage{color}
\input myunits.sty

\usepackage{amsmath}
\usepackage[T1]{fontenc}
\usepackage[english]{babel}
\usepackage{amsmath}
\usepackage{amsfonts}
\usepackage{mathrsfs}
\usepackage{amssymb}
\usepackage[ansinew]{inputenc}
\usepackage[font=small,format=plain,labelfont=bf,up]{caption}
\usepackage{hyperref}
\usepackage{cite}
\usepackage[dvipsnames]{xcolor}
\usepackage{pdfpages}
\usepackage{multirow}
\usepackage{array}
\usepackage{subcaption}
 
\begin{document}
\graphicspath{  {./figures/} }

\begin{titlepage}

\PHnumber{2016--206}
\PHdate{23 August 2016}

\title{Multiplicities of charged kaons from deep-inelastic muon scattering off an isoscalar target}

\Collaboration{The COMPASS Collaboration}
\ShortAuthor{The COMPASS Collaboration}

\begin{abstract}
Precise measurements of charged-kaon multiplicities in deep inelastic scattering were performed. The results are presented in three-dimensional bins
of the Bjorken scaling variable $x$, the relative virtual-photon energy $y$,
and the fraction $z$ of the virtual-photon energy carried by the produced hadron. The data were obtained by the COMPASS Collaboration by scattering $160\,\GeV$ muons off an isoscalar $^6$LiD target. They cover 
the kinematic domain $1 \,(\GeV/c)^2< Q^{2} < 60\,(\GeV/c)^2$ in the photon virtuality, $0.004 < x < 0.4$, $0.1 < y < 0.7$, $0.20 < z < 0.85$, and $W > 5\,\GeV/c^2$ in the invariant mass of the hadronic system. 
{The results from the sum of the $z$-integrated $\rm K^+$ and $\rm K^-$ multiplicities at high $x$ point to a value of the non strange quark fragmentation function larger than obtained by the earlier DSS fit.}
\end{abstract}

\vspace*{60pt}

\vfill
\Submitted{(to be submitted to Phys.\ Lett.\ B)}

\end{titlepage}

{\pagestyle{empty} 
%
%
\section*{The COMPASS Collaboration}
\label{app:collab}
\renewcommand\labelenumi{\textsuperscript{\theenumi}~}
\renewcommand\theenumi{\arabic{enumi}}
\begin{flushleft}
C.~Adolph\Irefn{erlangen},
J.~Agarwala\Irefn{calcutta},
M.~Aghasyan\Irefn{triest_i},
R.~Akhunzyanov\Irefn{dubna}, 
M.G.~Alexeev\Irefn{turin_u},
G.D.~Alexeev\Irefn{dubna}, 
A.~Amoroso\Irefnn{turin_u}{turin_i},
V.~Andrieux\Irefn{saclay},
N.V.~Anfimov\Irefn{dubna}, 
V.~Anosov\Irefn{dubna}
% ...
\end{flushleft}
%
%
\begin{Authlist}
\item \Idef{turin_p}{University of Eastern Piedmont, 15100 Alessandria, Italy}
\item \Idef{aveiro}{University of Aveiro, Department of Physics, 3810-193 Aveiro, Portugal}
\item \Idef{bochum}{Universit\"at Bochum, Institut f\"ur Experimentalphysik, 44780 Bochum, Germany\Arefs{l}\Arefs{s}}
\item \Idef{bonniskp}{Universit\"at Bonn, Helmholtz-Institut f\"ur  Strahlen- und Kernphysik, 53115 Bonn, Germany\Arefs{l}}
\item \Idef{bonnpi}{Universit\"at Bonn, Physikalisches Institut, 53115 Bonn, Germany\Arefs{l}}
\item \Idef{brno}{Institute of Scientific Instruments, AS CR, 61264 Brno, Czech Republic\Arefs{m}}
\item \Idef{calcutta}{Matrivani Institute of Experimental Research \& Education, Calcutta-700 030, India\Arefs{n}}
\item \Idef{dubna}{Joint Institute for Nuclear Research, 141980 Dubna, Moscow region, Russia\Arefs{o}}
\item \Idef{erlangen}{Universit\"at Erlangen--N\"urnberg, Physikalisches Institut, 91054 Erlangen, Germany\Arefs{l}}
\item \Idef{freiburg}{Universit\"at Freiburg, Physikalisches Institut, 79104 Freiburg, Germany\Arefs{l}\Arefs{s}}
\item \Idef{cern}{CERN, 1211 Geneva 23, Switzerland}
\item \Idef{liberec}{Technical University in Liberec, 46117 Liberec, Czech Republic\Arefs{m}}
\item \Idef{lisbon}{LIP, 1000-149 Lisbon, Portugal\Arefs{p}}
\item \Idef{mainz}{Universit\"at Mainz, Institut f\"ur Kernphysik, 55099 Mainz, Germany\Arefs{l}}
\item \Idef{miyazaki}{University of Miyazaki, Miyazaki 889-2192, Japan\Arefs{q}}
\item \Idef{moscowlpi}{Lebedev Physical Institute, 119991 Moscow, Russia}
\item \Idef{munichtu}{Technische Universit\"at M\"unchen, Physik Department, 85748 Garching, Germany\Arefs{l}\Arefs{r}}
\item \Idef{nagoya}{Nagoya University, 464 Nagoya, Japan\Arefs{q}}
\item \Idef{praguecu}{Charles University in Prague, Faculty of Mathematics and Physics, 18000 Prague, Czech Republic\Arefs{m}}
\item \Idef{praguectu}{Czech Technical University in Prague, 16636 Prague, Czech Republic\Arefs{m}}
\item \Idef{protvino}{State Scientific Center Institute for High Energy Physics of National Research Center `Kurchatov Institute', 142281 Protvino, Russia}
\item \Idef{saclay}{IRFU, CEA, Universit\'e Paris-Sud, 91191 Gif-sur-Yvette, France\Arefs{s}}
\item \Idef{taipei}{Academia Sinica, Institute of Physics, Taipei 11529, Taiwan}
\item \Idef{telaviv}{Tel Aviv University, School of Physics and Astronomy, 69978 Tel Aviv, Israel\Arefs{t}}
\item \Idef{triest_u}{University of Trieste, Department of Physics, 34127 Trieste, Italy}
\item \Idef{triest_i}{Trieste Section of INFN, 34127 Trieste, Italy}
\item \Idef{triest_ictp}{Abdus Salam ICTP, 34151 Trieste, Italy}
\item \Idef{turin_u}{University of Turin, Department of Physics, 10125 Turin, Italy}
\item \Idef{turin_i}{Torino Section of INFN, 10125 Turin, Italy}
\item \Idef{illinois}{University of Illinois at Urbana-Champaign, Department of Physics, Urbana, IL 61801-3080, USA}
\item \Idef{warsaw}{National Centre for Nuclear Research, 00-681 Warsaw, Poland\Arefs{u} }
\item \Idef{warsawu}{University of Warsaw, Faculty of Physics, 02-093 Warsaw, Poland\Arefs{u} }
\item \Idef{warsawtu}{Warsaw University of Technology, Institute of Radioelectronics, 00-665 Warsaw, Poland\Arefs{u} }
\item \Idef{yamagata}{Yamagata University, Yamagata 992-8510, Japan\Arefs{q} }
\end{Authlist}
%
%
\renewcommand\theenumi{\alph{enumi}}
\begin{Authlist}
\item [{\makebox[2mm][l]{\textsuperscript{\#}}}] Corresponding authors
\item \Adef{a}{Also at Instituto Superior T\'ecnico, Universidade de Lisboa, Lisbon, Portugal}
\item \Adef{b}{Also at Department of Physics, Pusan National University, Busan 609-735, Republic of Korea and at Physics Department, Brookhaven National Laboratory, Upton, NY 11973, USA}
\item \Adef{r}{Supported by the DFG cluster of excellence `Origin and Structure of the Universe' (www.universe-cluster.de)}
\item \Adef{d}{Also at Chubu University, Kasugai, Aichi 487-8501, Japan\Arefs{q}}
\item \Adef{x}{Also at Department of Physics, National Central University, 300 Jhongda Road, Jhongli 32001, Taiwan}
\item \Adef{e}{Also at KEK, 1-1 Oho, Tsukuba, Ibaraki 305-0801, Japan}
\item \Adef{g}{Also at Moscow Institute of Physics and Technology, Moscow Region, 141700, Russia}
\item \Adef{v}{Supported by Presidential grant NSh--999.2014.2}
\item \Adef{h}{Present address: RWTH Aachen University, III.\ Physikalisches Institut, 52056 Aachen, Germany}
\item \Adef{y}{Also at Department of Physics, National Kaohsiung Normal University, Kaohsiung County 824, Taiwan}
\item \Adef{i}{Present address: Uppsala University, Box 516, 75120 Uppsala, Sweden}
\item \Adef{c}{Supported by the DFG Research Training Group Programme 1102  ``Physics at Hadron Accelerators''}
%
%
\item \Adef{l}{Supported by the German Bundesministerium f\"ur Bildung und Forschung}
\item \Adef{s}{Supported by EU FP7 (HadronPhysics3, Grant Agreement number 283286)}
\item \Adef{m}{Supported by Czech Republic MEYS Grant LG13031}
\item \Adef{n}{Supported by SAIL (CSR), Govt.\ of India}
\item \Adef{o}{Supported by CERN-RFBR Grant 12-02-91500}
\item \Adef{p}{\raggedright Supported by the Portuguese FCT - Funda\c{c}\~{a}o para a Ci\^{e}ncia e Tecnologia, COMPETE and QREN,
 Grants CERN/FP 109323/2009, 116376/2010, 123600/2011 and CERN/FIS-NUC/0017/2015}
\item \Adef{q}{Supported by the MEXT and the JSPS under the Grants No.18002006, No.20540299 and No.18540281; Daiko Foundation and Yamada Foundation}
\item \Adef{t}{Supported by the Israel Academy of Sciences and Humanities}
\item \Adef{u}{Supported by the Polish NCN Grant 2015/18/M/ST2/00550}
\end{Authlist}
}
\newpage
\setcounter{page}{1}
\parindent=0em

\section{Introduction}
\vspace{0.2cm}
The study of strangeness in the nucleon has been arousing constant interest over the last decades, but the data sensitive to strangeness are still rather scarce. The parton distribution functions (PDFs) of strange quarks $s(x)$ in the nucleon are not yet well constrained. The evaluation of the s-quark helicity distribution function $\Delta s(x)$, where $x$ denotes the fraction of longitudinal nucleon momentum carried by the quark, is particularly difficult. Different values are obtained for its first moment $\Delta s$ when using two different methods. On the one hand, a substantially negative value is obtained from {\it inclusive} measurements of deep-inelastic scattering (DIS) of a polarised beam on a polarised target, when assuming SU(3) flavour symmetry in the hyperon sector (see for instance Ref.~\citen{COMPASSg1}).
On the other hand, the values of $\Delta s(x)$ obtained from {\it semi-inclusive} DIS measurements (hereafter referred to as SIDIS) of identified kaons using polarised beam and target are found to be compatible with zero in the measured kinematic range~\cite{COMPASSsidis, COMPASSasym,HERMESdeltas}. The latter evaluations require knowledge of the collinear quark fragmentation functions (FFs), in particular knowledge of the quark-to-kaon FF.
The values of the strange quark-to-kaon FF presently published 
 differ by factors of up to three~\cite{EMCFFs, DSSFFs} and thus can lead to very different values of $\Delta s(x)$~\cite{COMPASSasym, LSS}. 
Quark FFs are of general interest by themselves, as they are non-perturbative universal functions describing processes leading to the production of hadrons. 

The reason for the presently large uncertainties on the values of the strange quark-to-kaon FF is the lack of appropriate kaon data. Existing data on kaon production from e$^+$e$^-$ annihilation~\cite{epem_data} and pp collision~\cite{RHIC} experiments are 
only weakly sensitive to the quark and anti-quark flavours individually. In contrast, SIDIS data allow for a full flavour decomposition of FFs~\cite{LSS-FF}. Compared to the kinematic range of the only published SIDIS results, which originate from HERMES~\cite{HERMESkaons}, COMPASS measurements cover a wider kinematic domain. In addition, they better satisfy the Berger criterion~\cite{berger}
that ensures the validity of factorisation in the description of SIDIS.  Kaon multiplicities measured by COMPASS are thus expected to play an important role in future next-to-leading order (NLO) pQCD analyses aiming to constrain quark-to-kaon FFs. 

In this Letter, we present COMPASS results on differential multiplicities for charged kaons, which are derived from data taken simultaneously to those used for the determination of the differential multiplicities for charged pions~\cite{pionpaper}. In both cases, similar analysis methods are used.
The multiplicity results are derived from
SIDIS measurements with polarised beam and target after averaging over the target polarisation, so that they cover the same kinematic range as used for the extraction of $\Delta s(x)$~\cite{COMPASSasym}, 
as well as a large $Q^2$ range to probe evolution.

The multiplicity for a hadron of type $\rm h$ from the SIDIS measurement lN $\rightarrow$ lhX is defined as the differential hadron production cross section $\sigma^{\rm h}$
normalised to the inclusive DIS cross section $\sigma^{\rm DIS}$ and thus represents the mean number of hadrons produced per DIS event when integrated over $z$:
\begin{equation}
\frac{{\rm d}M^{\rm h}(x,z,Q^2)}{{\rm d}z}=
\frac{{\rm d}^3\sigma^{\rm h}(x,z,Q^2)/{\rm d}x{\rm d}Q^2{\rm d}z}{{\rm d}^2\sigma^{\rm DIS}/{\rm d}x{\rm d}Q^2}.
\label{MulDef}
\end{equation}
Here,
$x = -q^{2}/(2P\cdot q)$ is the Bjorken variable, $z = (P\cdot p_{\rm h})/(P\cdot q)$ the fraction of the 
virtual-photon energy that is carried by the final-state hadron and $Q^{2}= - {q}^{2}$ the negative square of the lepton four-momentum transfer. 
The symbols $q$, $P$ and $p_{\rm h}$ denote the four-momenta of the virtual photon, the 
nucleon N and the observed hadron h respectively.
Additional variables used are the lepton energy fraction carried by the virtual 
photon, {$y=(P\cdot q)/(P\cdot k)$ 
with $k$ the four-momentum of the incident lepton,}
and the invariant mass of the final hadronic system, $W = \sqrt{(P + q)^{2}}$. 
At LO in pQCD the cross sections are not sensitive to the gluon contribution and can be expressed in terms of only quark PDFs and FFs:

\begin{equation}
\frac{{\rm d}^2\sigma^{\rm DIS}}{{\rm d}x{\rm d}Q^2}=C(x,Q^2)\sum_{\rm q}e_{\rm q}^2q(x,Q^2), \qquad
\frac{{\rm d}^3\sigma^{\rm h}}{{\rm d}x{\rm d}Q^2{\rm d}z}=C(x,Q^2)\sum_{\rm q}e_{\rm q}^2q(x,Q^2)D_{\rm q}^{\rm h}(z,Q^2).
\label{Sigma}
\end{equation}
Here, $q(x,Q^2)$ is the quark PDF for flavour $\rm q$, $D_{\rm q}^{\rm h}(z,Q^2)$ denotes the FF that describes the fragmentation of a quark of flavour $\rm q$ to a hadron of type $\rm h$, 
$C(x,Q^2)={2\pi\alpha^2(1+(1-y)^2)}/{Q^4}$ and $\alpha$ is the fine structure constant. 

\section{Experimental setup and data analysis}\label{analysis}
\vspace{0.2cm}
The data were taken in 2006 using muons from the M2 beam line of the CERN SPS. 
The beam momentum was $160\,\GeV/c$ with a spread of $\pm$ 5\%. 
The solid-state $^6$LiD target was effectively isoscalar, with
a small excess of neutrons over protons of about 0.2\% due to 
the presence of additional material in the target (H, $^3$He and $^7$Li). 
The target was longitudinally polarised but in the present analysis
the data are averaged over the target polarisation. 
The experimental setup used the same two-stage COMPASS spectrometer~\cite{Abbon:2007pq} as described in the recent publication on the pion multiplicity measurement~\cite{pionpaper}.
For pion, kaon and proton 
separation, the ring imaging Cerenkov 
counter (RICH) was used.
It was filled with C$_4$F$_{10}$ radiator gas leading to thresholds for
pion, kaon and proton detection of about $2.9\,\GeV/c$, $9\,\GeV/c$ and $18\,\GeV/c$ 
respectively. The central region from the RICH is excluded in the analysis in order to avoid particles crossing the beam pipe inside the RICH.

The data analysis includes event selection, particle 
identification and acceptance correction, as well as corrections for QED radiative effects and
diffractive vector-meson production. 
The kaon multiplicities \mbox{$M^{\rm K}$($x$, $y$, $z$)}  are determined from the 
kaon yields $N^{\rm K}$ normalised by the number of DIS events, $N^{\rm DIS}$, and divided by the acceptance correction $A$:
\begin{equation}
\label{ExpMul}
\frac{\text{d}M^{\rm K}(x,y,z)}{\text{d}z} = 
\frac{1}{N^{\rm DIS}(x,y)}\frac{\text{d}N^{\rm K}(x,y,z)}{\text{d}z} \frac{1}{A(x,y,z)}\,.
\end{equation}
Here, $y$ is used as a third variable instead of $Q^2$ because of the strong correlation between $x$ and $Q^2$ in the COMPASS fixed-target kinematics. 

\subsection{Event and hadron selection}
\label{sec:event}
The present analysis is based on events with inclusive triggers that only use information on the scattered muon.
A total of 13~$\times~10^{6}$ DIS events were collected for an integrated luminosity of 0.54~fb$^{-1}$. The data cover a wide kinematic range of $1\,(\GeV/c)^2 < Q^{2} < 60\,(\GeV/c)^2$, $0.004<x<0.4$ and  $W>5\,\GeV/c^2$.
The relative virtual-photon energy $y$ is constrained to the range $0.1 < y < 0.7$ in order to exclude kinematic regions where the momentum resolution degrades and radiative effects are most pronounced. Further constraints on $y$ are discussed in Sect.~\ref{sec:acceptance}.
For the selection of kaons in the SIDIS final state, $z$ is constrained to $0.2 \le z \le 0.85$. The lower limit avoids the contamination 
from target remnant fragmentation, while the upper limit excludes muons wrongly 
identified as hadrons. Further constraints on momentum and 
polar angle of hadrons are discussed in Sect.~\ref{RICH}.

The corrections for higher-order QED effects are calculated on an
event-by-event basis and depend on $x$ and $y$ at a given beam energy. The
correction factors are calculated separately for denominator and numerator of
Eq.~\ref{ExpMul}, i.e.\ for the number of DIS events and
the number of kaons, respectively. The Dubna TERAD code is used~\cite{bardin} for the DIS correction factor.
For the kaon yields, the TERAD correction is calculated excluding the elastic and quasi-elastic tails. As TERAD cannot account for a $z$ dependence, which leads to an overestimate of
the correction,\footnote{We also estimated the radiative corrections using the RADGEN code~\cite{RADGEN} together with the LEPTO generator, which effectively incorporates a $z$ dependence. However, by including RADGEN and LEPTO in the COMPASS full MC chain we were not able to correctly describe the data. Note also that an author of RADGEN advises caution when
using the program to correct differential cross section
distributions for radiative effects beyond the (x,y) dependence~\cite{JLabrad_conf}.}
a conservative approach is adopted here. The correction is calculated for the two extreme cases, no correction and full
correction to the number of kaons entering the numerator; half of the correction is applied to the multiplicities. This approach leads to an overall correction between $1\%$ and $7\%$ depending on kinematics.

%

\subsection{Kaon identification using the RICH detector}\label{RICH}
\vspace{0.2cm}
The hadron identification using the RICH detector~\cite{Abbon:2011zza} 
and the unfolding procedure for identified hadrons of different types follow the method described in Ref.~\citen{pionpaper}. Only the main features are recalled here for completeness.
A maximum likelihood method is used, based
on the pattern of photons
detected in the RICH detector. The likelihood values are calculated by comparing 
the measured photo-electron pattern with those expected
for different mass hypotheses ($\pi$, K, p), taking the distribution of 
background photons into account. 

\begin{figure}[htbp]
\centering
\includegraphics[width=.55\textwidth]{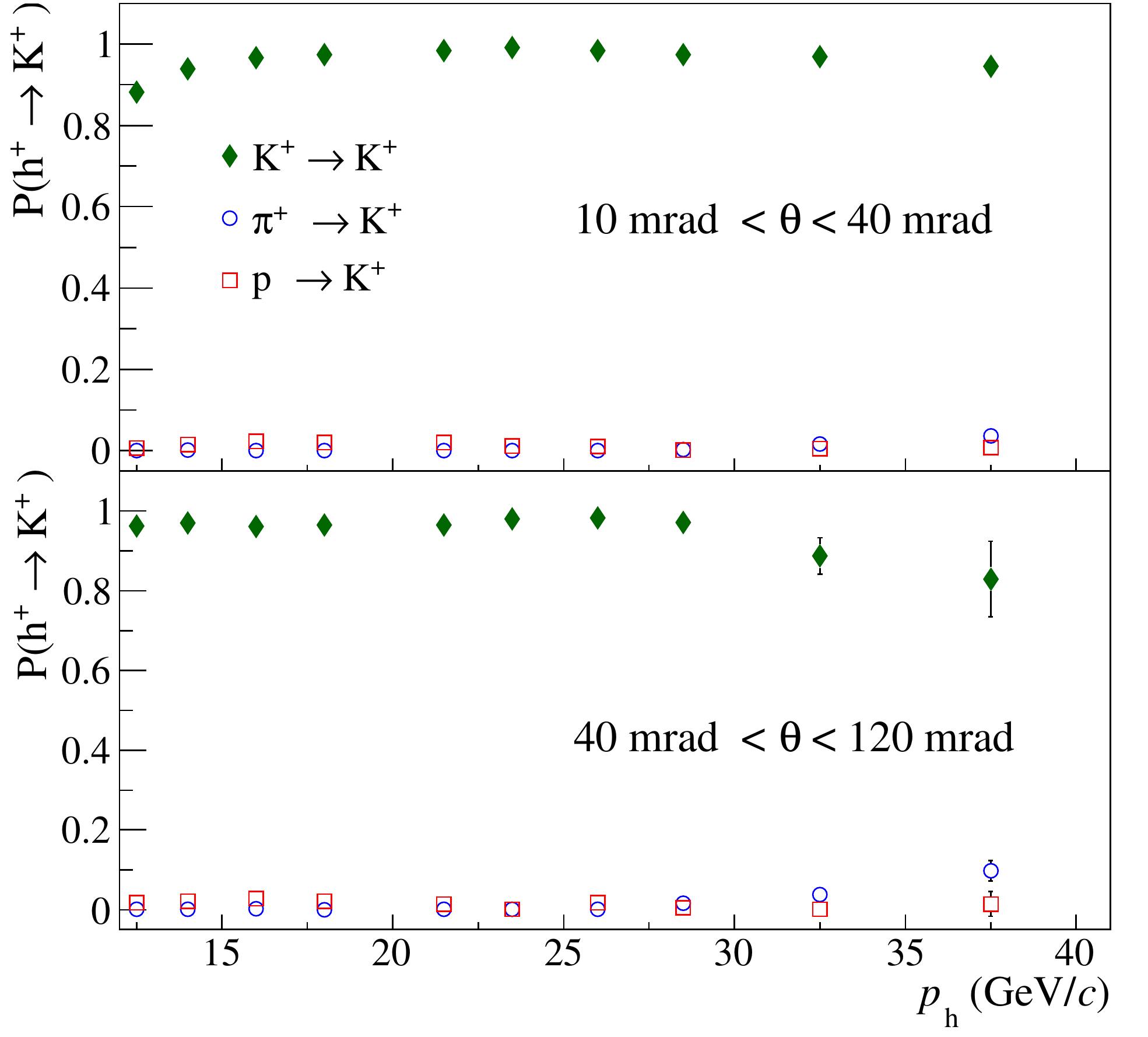}
\caption{Probabilities of RICH identification of $\pi^+$, K$^+$ and p as ${\rm K}^+$, shown versus momentum for the two $\theta$ regions. 
}
\label{RichEff}
\end{figure}

The yields of identified hadrons, $N_{\rm true}$, are obtained by applying an unfolding algorithm 
to the yields of measured hadrons, $N_{\rm meas}$, in order to correct for the identification and misidentification probabilities:
\begin{equation}
N^{i}_{\rm true} = \sum_{j} (\text{P}^{-1})_{ij} \cdot N^{j}_{\rm meas}\,.
\label{Relation1}
\end{equation}
These probabilities are calculated in a matrix $ {\rm P}_{ij}$ 
with 
$i,j \,\ \epsilon  \{ \pi, {\rm K}, {\rm p}\}$, which contains as diagonal elements  the efficiencies and 
as off-diagonal 
elements the misidentification probabilities. 
The elements of this $3 \times 3$ matrix are constrained by 
$\sum_{j} {\rm P}_{ij} \le 1$. 
They are determined from data using samples of $\pi$, K and p that originate from the decay of K$^0_{\rm S}$, $\phi$, and $\Lambda$ respectively, into two charged 
particles.
The probabilities depend mostly on particle momentum $p_{\rm h}$ and polar angle $\theta$ at the entrance of the RICH detector.
The $p_{\rm h}$ dependence accounts for effects arising from momentum 
thresholds (see the beginning of Sect.\ref{analysis}) and from saturation at high momentum
($\beta \rightarrow 1$). The $\theta$ dependence that accounts for 
varying 
occupancy and background levels in the RICH photon detectors is weak, so that it is sufficient to divide in two the full polar angle region where 
the efficiencies are high and precisely measured ($10\,\mrad < \theta < 40\,\mrad$ and $40\,\mrad < \theta < 120\,\mrad$). 
In order to achieve excellent kaon-pion 
separation and high particle-identification probabilities for kaons,
only particles with momenta between $12\,\GeV/c$ and $ 40\,\GeV/c$ are selected.
Figure~\ref{RichEff} shows the probabilities of positive hadrons to be identified as 
${\rm K}^+$ vs momentum $p_{\rm h}$ for the two $\theta$ regions.
Kaons are identified with a 95\% efficiency over most of the momentum range with probabilities to misidentify pions and protons as kaons being smaller than 3\%. 
Similar values are
obtained in the case of negative charge. 
The number $N^{{\rm K}^\pm}_{\rm true}$ of identified kaons 
available for the analysis after all 
particle identification cuts is 
2.8~$\times~10^{6}$. 

\subsection{Acceptance correction}\label{sec:acceptance}
The overall correction for the geometric and 
kinematic acceptance of the COMPASS apparatus and for detector
inefficiencies, resolutions and bin migration is evaluated using a Monte Carlo (MC) simulation. A good description of the kinematic distributions of experimental data is obtained using the following tools for the MC simulation: 
LEPTO \cite{Ingelman:1996mq} for the generation of DIS events, JETSET~\cite{Sjostrand:1995iq} for hadronisation
(tuning from Ref.~\citen{Adolph:2012vj}), 
GEANT3~\cite{Brun:1985ps} for the description of the spectrometer
and FLUKA~\cite{FLUKA} for secondary hadron interactions. 
The acceptance correction is calculated in narrow regions of \mbox{($x$, $y$, $z$)}, thus reducing the dependence on the choice of the generator used in the simulation. It is defined as ratio of reconstructed 
and generated multiplicities: 
\begin{equation}
     A(x ,y ,z) 
 = \frac { {\rm d} N ^{\text{K}} _{\text{rec}}(x,y,z) / N ^{\text{DIS}} _{\text{rec}}(x,y) } {{\rm d} N^{\text{K}}_{\text{gen}}(x,y,z) /N^{\text{DIS}}_{\text{gen}}(x,y)}.
\label{accep}
 \end{equation}
The generated kinematic variables are used for the generated events, while 
the reconstructed kinematic variables are used for the reconstructed events.
The reconstructed MC hadrons are subject 
to the same kinematic and geometric selection criteria as the data,
while the generated hadrons are subject to 
kinematic requirements only. 
The average value of the acceptance correction is about 65\%.
The acceptance correction is almost flat in $z$, $x$ and $y$, except at high $x$ and low $y$, but it is always larger than 35\%.
For each bin in $z$, the 
$y$ range for DIS events is restricted to the region accessible with kaon momenta within the range from $12\,\GeV/c$ to $40\,\GeV/c$.

\subsection{Vector meson correction} 
\label{sect:VM}
For both the DIS and the kaon SIDIS samples, it is necessary to estimate the fraction of events, in which diffractively produced vector mesons have decayed into lighter hadrons. This fraction can be considered as a higher-twist contribution to the SIDIS cross section~\cite{HERMES2008}. It cannot be described by the pQCD parton model with the independent-fragmentation mechanism that is encoded in the FFs. Hence FFs extracted from data including this fraction would be biased and in particular the universality principle of the model would be violated.
The correction for diffractive vector meson decay is applied separately in the DIS and kaon samples.
Monte Carlo SIDIS events that are free of diffractive contributions are simulated using LEPTO. Diffractive production of exclusive $\rho^0$ and 
$\phi$ events is simulated using HEPGEN~\cite{hepgen}, while further exclusive-meson production channels characterised
by smaller cross sections and more particles in the final state are not
taken into account. Vector meson production with diffractive
dissociation of the target nucleon is also simulated and represents about 20\% of the
cross section given by HEPGEN. 

\begin{figure}[htbp]
\centering
\includegraphics[width=.55\textwidth]{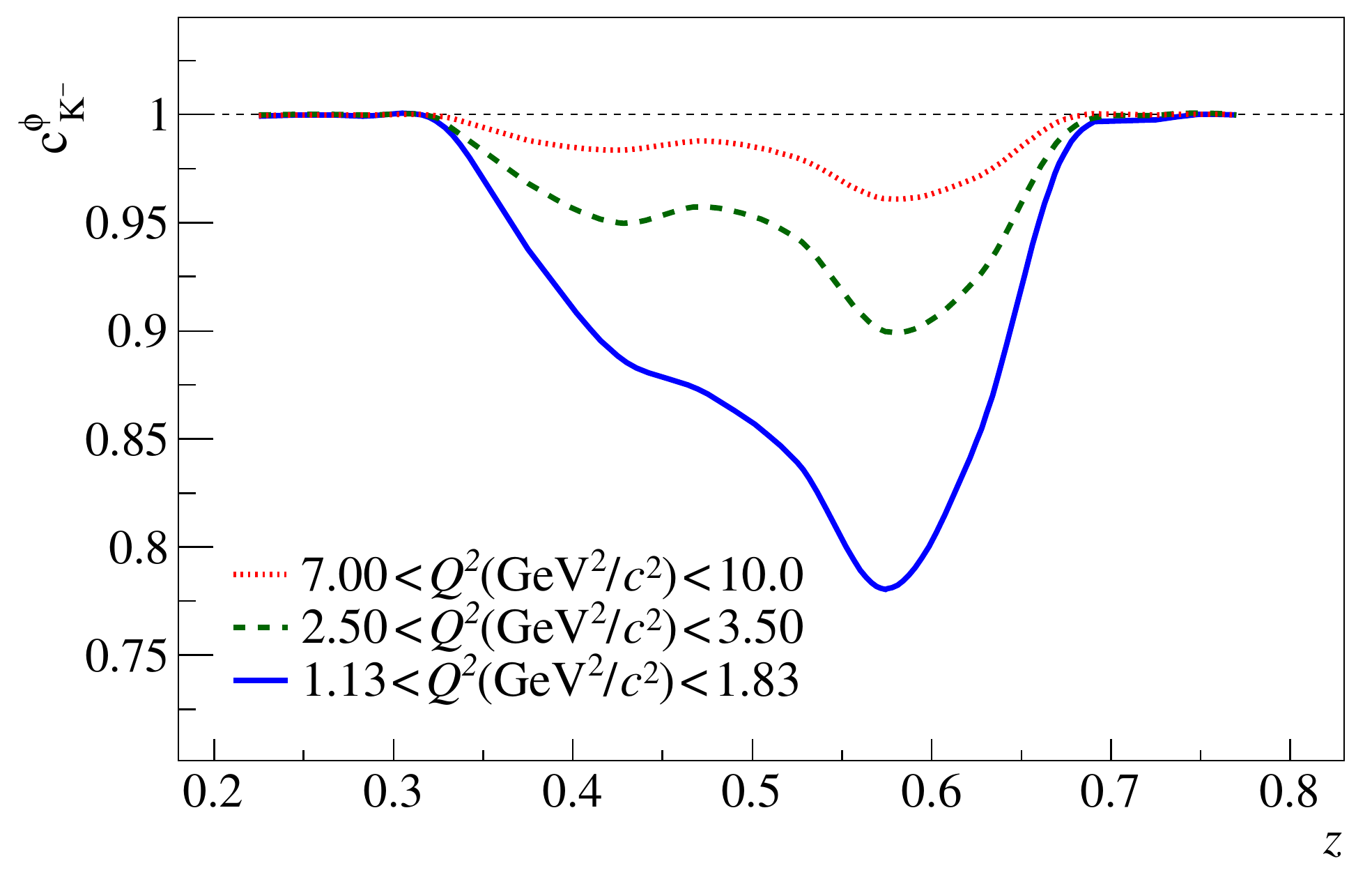}
\caption{Correction factor for negative kaon multiplicities due to diffractive production of vector meson $\phi$ shown versus $z$ for three $Q^{2}$ regions. 
}
\label{DVM11}
\end{figure}

\begin{figure}[htbp]
\centering
\includegraphics[width=.55\textwidth]{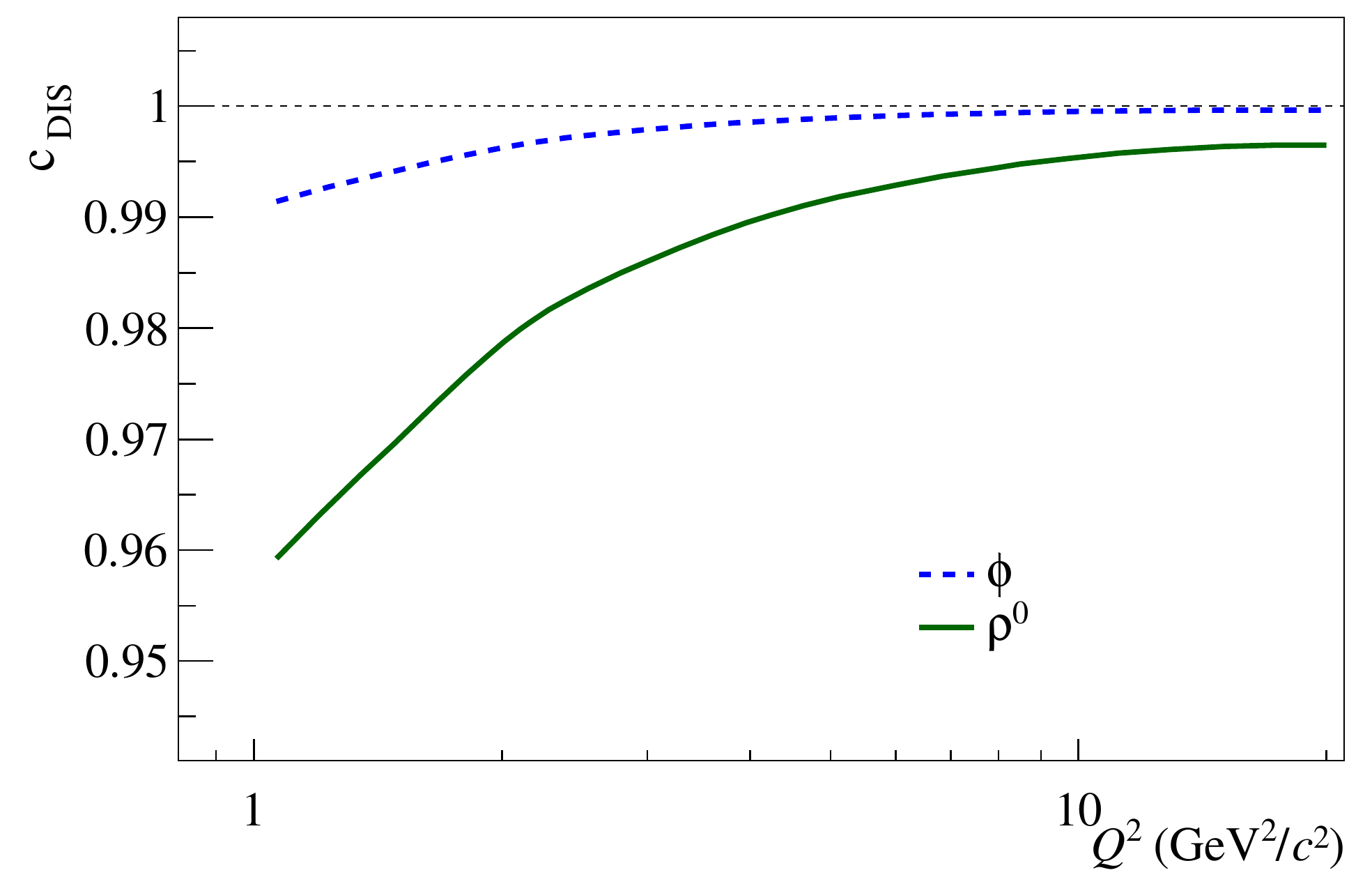}
\caption{Correction factor for DIS events due to diffractive production of vector mesons $\phi$ and $\rho^0$ shown versus $Q^{2}$. 
} 
\label{DVM2}
\end{figure}

An example of the correction factor calculated for the diffractive contribution of $\phi$-meson decay into kaons, $\rm c^{\rm \phi}_{\rm K^{-}}$ vs $z$, is shown in Fig.\,\ref{DVM11} for three different $Q^2$ ranges. It varies between $0.1\%$ and $22\%$ with the largest contribution appearing at small $Q^2$
and $z$ close to $0.6$. The correction to DIS yields, $\rm c_{\rm DIS}$ vs $Q^2$, is shown in Fig.\,\ref{DVM2}. It is of the order of a few percent.


\subsection{Systematic uncertainties}

\vspace{0.2cm}
For the evaluation of systematic uncertainties the same studies are repeated as
they were performed for the analysis of pions~\cite{pionpaper} and summarised 
here. A maximum uncertainty of $4\%$ on the acceptance correction is 
derived from the following studies: varying the PDF set used, varying the JETSET parameters, using a multi-dimensional space with additional variables and varying the detector efficiency applied in the MC simulation used for the acceptance calculation, as well as comparing multiplicity results from upstream and downstream target regions. The RICH identification and unfolding procedure leads to uncertainties typically below $1.5\%$. The uncertainty on the diffractive meson correction factor is on average below $1.5\%$; it is estimated assuming $30\%$ uncertainty on the cross section for exclusive production of $\phi$ mesons~\cite{GK}, which is used to normalise the HEPGEN MC calculation. Possible nuclear effects due to the presence of $^3$He/$^4$He and $^6$Li in the target material are expected to be very small and hence neglected. No time dependence is observed in the results obtained in $6$ weeks of data taking. Also, the present results are found in agreement with an independent set of preliminary data taken in 2004 before the upgrade of the spectrometer. We chose to publish only the present set of data, which is of better quality and benefits from a larger spectrometer acceptance and an upgraded RICH detector.

In addition, a conservative estimate of $100\%$ systematic uncertainty is assumed for the corrections for higher-order QED effects. This corresponds to a $1\%$ to $7\%$ systematic uncertainty on the multiplicities.

All these contributions are added in quadrature and yield a total systematic uncertainty, $\sigma_{syst}$, which varies between $5\%$ and $10\%$ depending on kinematics. It is generally larger than the statistical one at low $z$, whereas at higher $z$ the statistical uncertainty dominates. 

From the total systematic uncertainty a large fraction,
$0.8 \sigma_{syst}$, is estimated to be correlated from bin to bin, and the remaining fraction, $\sqrt{(1-0.8^2)} \sigma_{syst}$, is uncorrelated.

\newcommand{\fav}{\text{fav}}
\newcommand{\unf}{\text{unf}}
\newcommand{\soutr}{\text{str}}
\newcommand{\Dfav}{D_{\fav}}
\newcommand{\Dunf}{D_{\unf}}
\newcommand{\Dstr}{D_{\soutr}}
\newcommand{\Dg}{D_{\g}}
\newcommand{\ubar}{\overline{\text{u}}}
\newcommand{\dbar}{\overline{\text{d}}}
\newcommand{\sbar}{\overline{\text{s}}}
\newcommand{\normDIS}{5(u + d + \ubar + \dbar) + 2(s + \sbar)}
\section{Results for charged-kaon multiplicities}\label{sec:results}
\begin{figure}[htbp]
\centering
\includegraphics[width=\hsize]{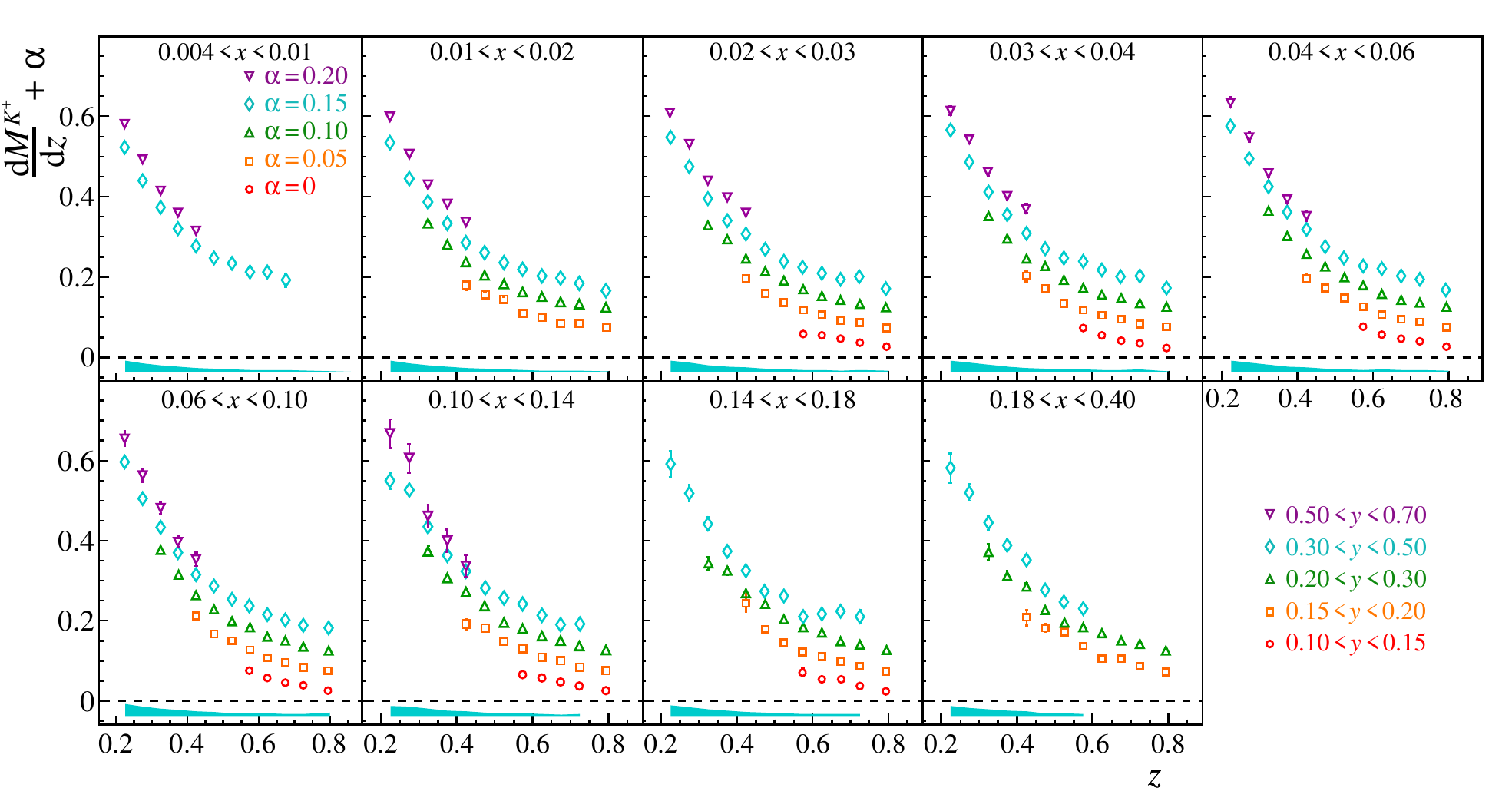}
\caption{Multiplicities for $K^+$ shown vs $z$ for the nine $x$ ranges and the five 
$y$ ranges (for better visibility staggered vertically by a constant $\alpha$). The band corresponds to the total systematic uncertainty for the range $0.30<y<0.50$.
}
\label{res1}
\end{figure}

\begin{figure}[htbp]
\centering
\includegraphics[width=\hsize]{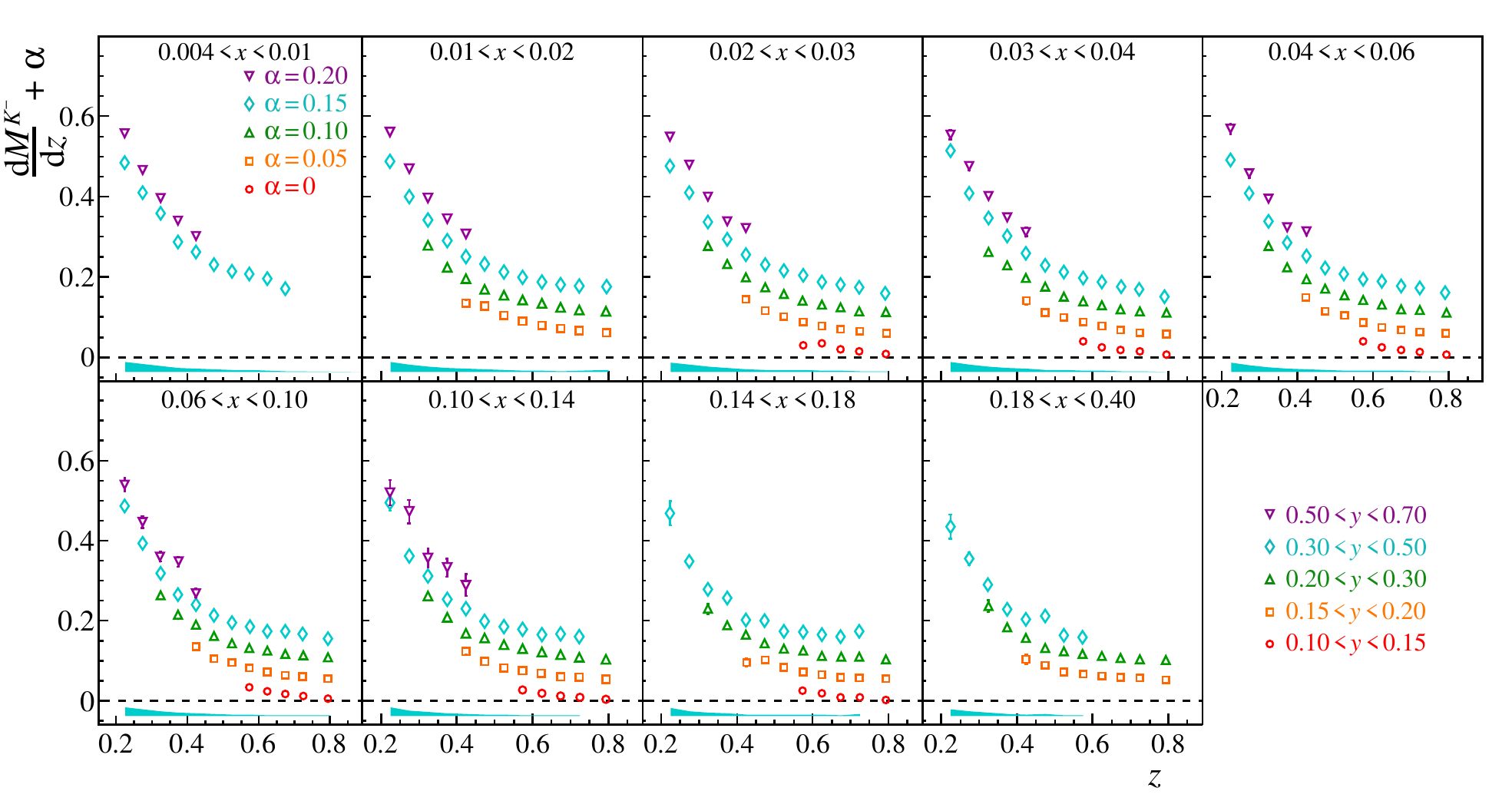}
\caption{Same as Fig.~\ref{res1} but for negative kaons. 
} 
\label{res2}
\end{figure}

\begin{figure}[!h]
\centering
\includegraphics[width=\hsize]{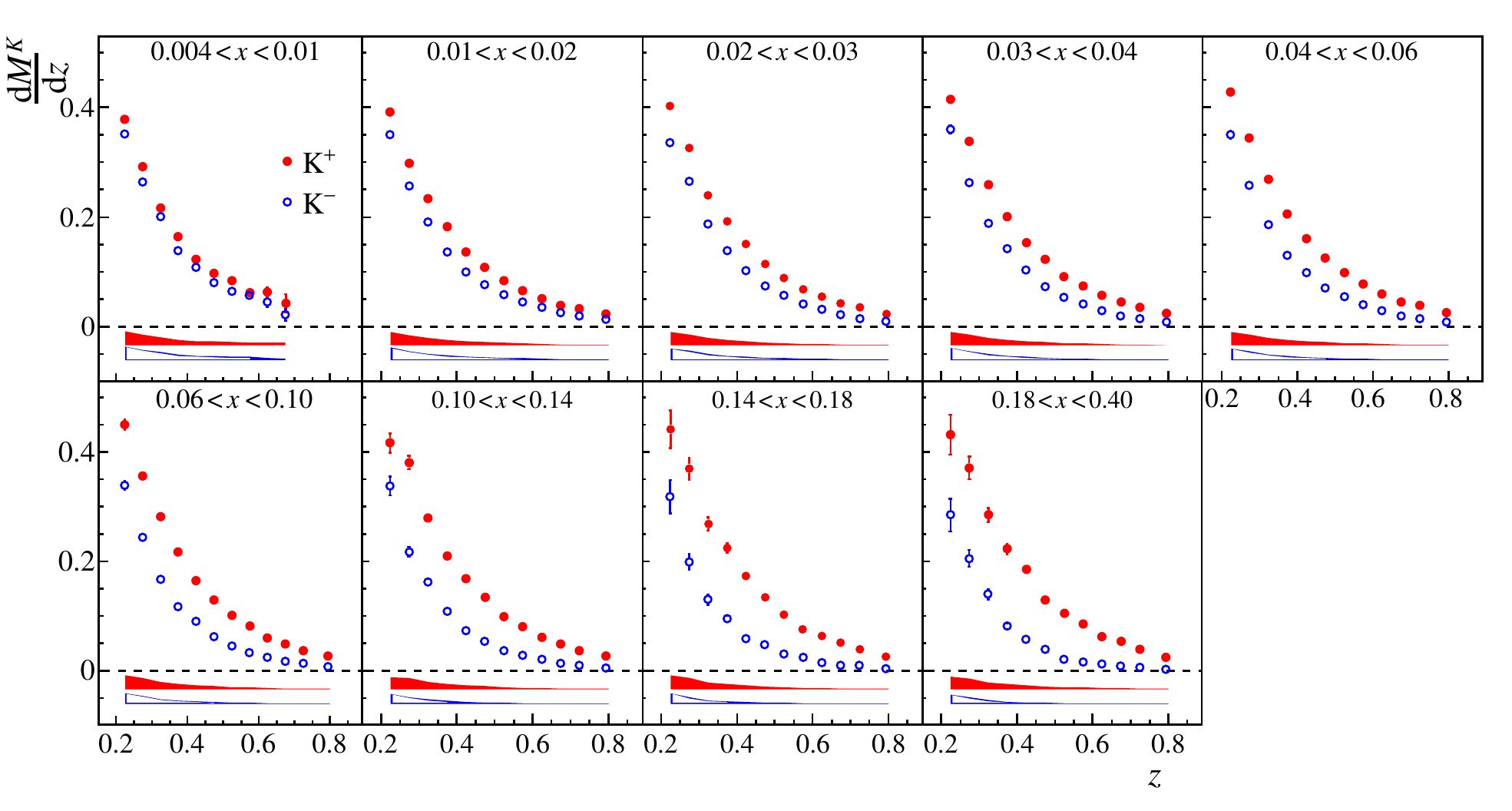}
 \caption{Positive (closed) and negative (open) kaon multiplicities vs $z$ for the 
nine $x$ ranges. The bands correspond to the total systematic uncertainties. 
}
 \label{res3}
\end{figure}

The results for $\rm K^+$ and $\rm K^-$ multiplicities represent a total of 
$630$ experimental data points. In this section, results are shown first in the
initial three-dimensional ($x$, $y$, $z$) representation, then averaged in $y$,
and eventually also integrated over the $z$ range from $0.2$ to $0.85$ in order
to calculate the sum and ratio of $\rm K^+$ and $\rm K^-$ multiplicities vs 
$x$. The data presented in the following figures are all corrected for
the diffractive vector-meson contribution as described in Sect.~\ref{sect:VM}.
For completeness, the numerical values are available on HepData~\cite{HEPDATA} 
for multiplicities with and without this correction.
The separate correction factors for DIS and kaon yields are also provided, as 
are the factors used for the correction of radiative effects. The bin limits 
in $x$, $y$, and $z$ are given in Table~\ref{tab:Binning}. 
 \begin{table}[!htbp]
  \caption{Bin limits for the three-dimensional ($x$, $y$, $z$) representation.}
 \centering
    \begin{tabular}{|l|rrrrrrrrrrrrr|}
      \hline
     &\multicolumn{13}{|c|} {bin limits}\\
      \hline
      $x$  &0.004 &0.01& 0.02&0.03&0.04&0.06&0.1&0.14&0.18&0.4&&&\\
     \hline
      $y$ & 0.1&0.15&0.2&0.3&0.5&0.7&&&&&&&\\
     \hline
      $z$ & 0.2&0.25&0.3&0.35&0.4& 0.45&0.5&0.55&0.6&0.65&0.7&0.75&0.85\\
      \hline
     \end{tabular}
  \label{tab:Binning}
\end{table}
The $Q^2$ values range from $1\,(\GeV/c)^2$ at 
the smallest $x$ to about $60\,(\GeV/c)^2$ at the largest $x$, 
with $\langle Q^2 \rangle = 3\,(\GeV/c)^2$. In Fig.~\ref{res1} and Fig.~\ref{res2}, the results for the $z$ and $y$ dependences of the $\rm K ^+$ and $\rm K^-$
multiplicities are presented separately for the nine ranges in $x$. Statistical uncertainties 
are shown for all points, while the band on the bottom of the plots illustrates the size of the total systematic uncertainty for a single $y$ range ($0.30 - 0.50$). For the other $y$ ranges, the bands are similar or smaller in size. 
In Fig.~\ref{res3}, multiplicities of positive (closed circles) and 
negative (open circles) charged kaons are shown vs $z$, separately for 
the nine $x$ ranges but averaged over $y$. The error bars correspond to the 
statistical uncertainties and the bands to the total systematic ones. 
A strong dependence on $z$ is observed as in previous measurements~\cite{HERMES2013} and a weak one on $x$. The multiplicities are systematically higher for $\rm K^+$ as compared to $\rm K^-$ because of $u$-quark dominance and since $\rm K^-$ do not contain valence quarks from the nucleon.

\begin{figure}[!h]
\centering
\includegraphics[width=\hsize]{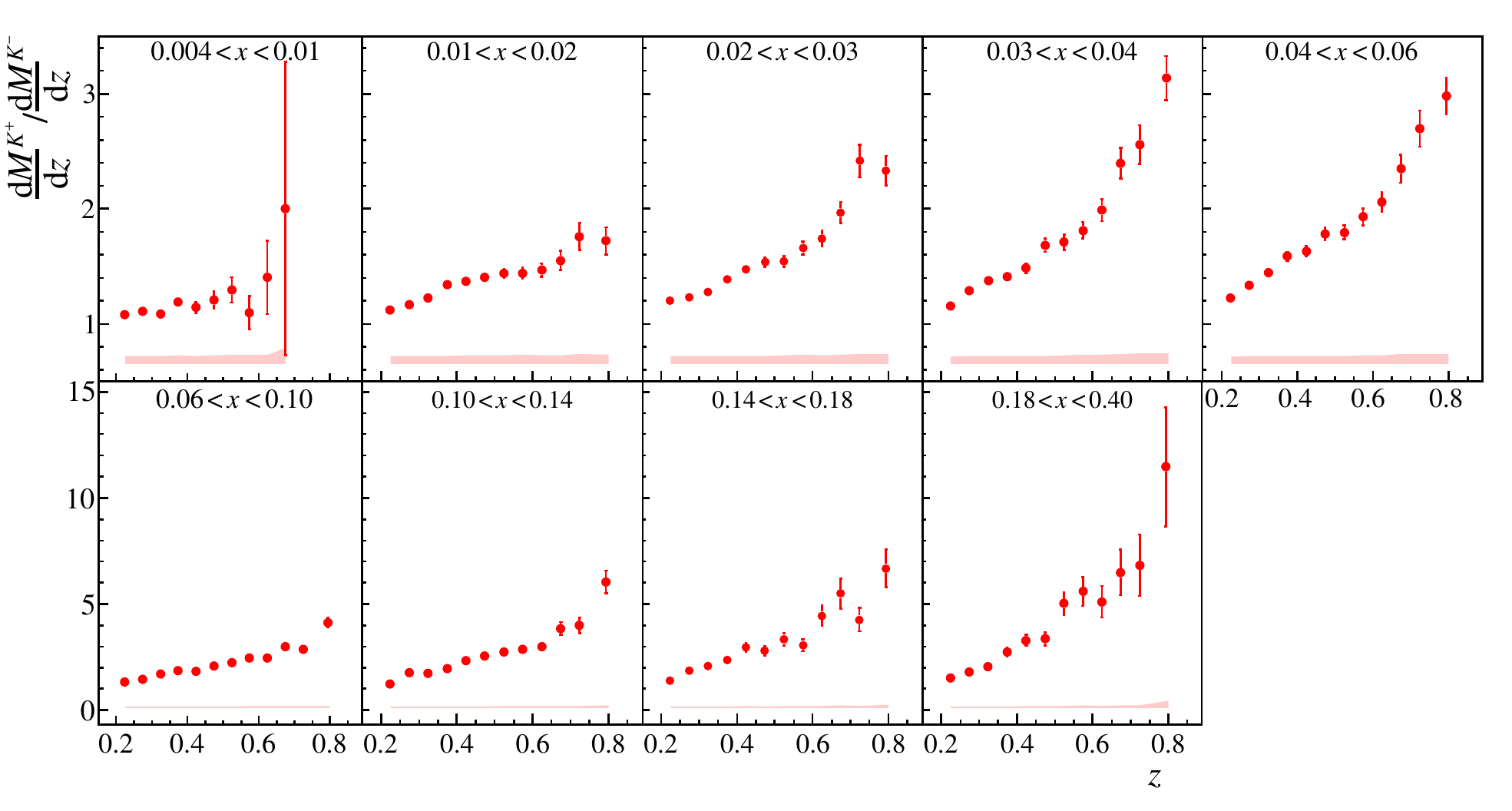}
 \caption{Ratio of multiplicities $($d$M^{\rm K^+}/$d$z) / ($d$M^{\rm K^-}/$d$z)$ vs $z$ for the nine $x$ ranges. The bands correspond to the total systematic uncertainties. 
 }
 \label{fig:ratio3d}
\end{figure}

%
%
In Fig.~\ref{fig:ratio3d}, we present the ratio $R^{\rm K^+/\rm K^-} = ($d$M^{\rm K^+}/$d$z) / ($d$M^{\rm K^-}/$d$z$) vs $z$ in the nine $x$ ranges. In this ratio, most of the systematic uncertainties cancel, including the one on radiative corrections, so that it provides a benchmark for future NLO fits. A steep rise in $z$ is observed.

\begin{figure}[htbp]
\centering
\includegraphics[width=8.6cm]{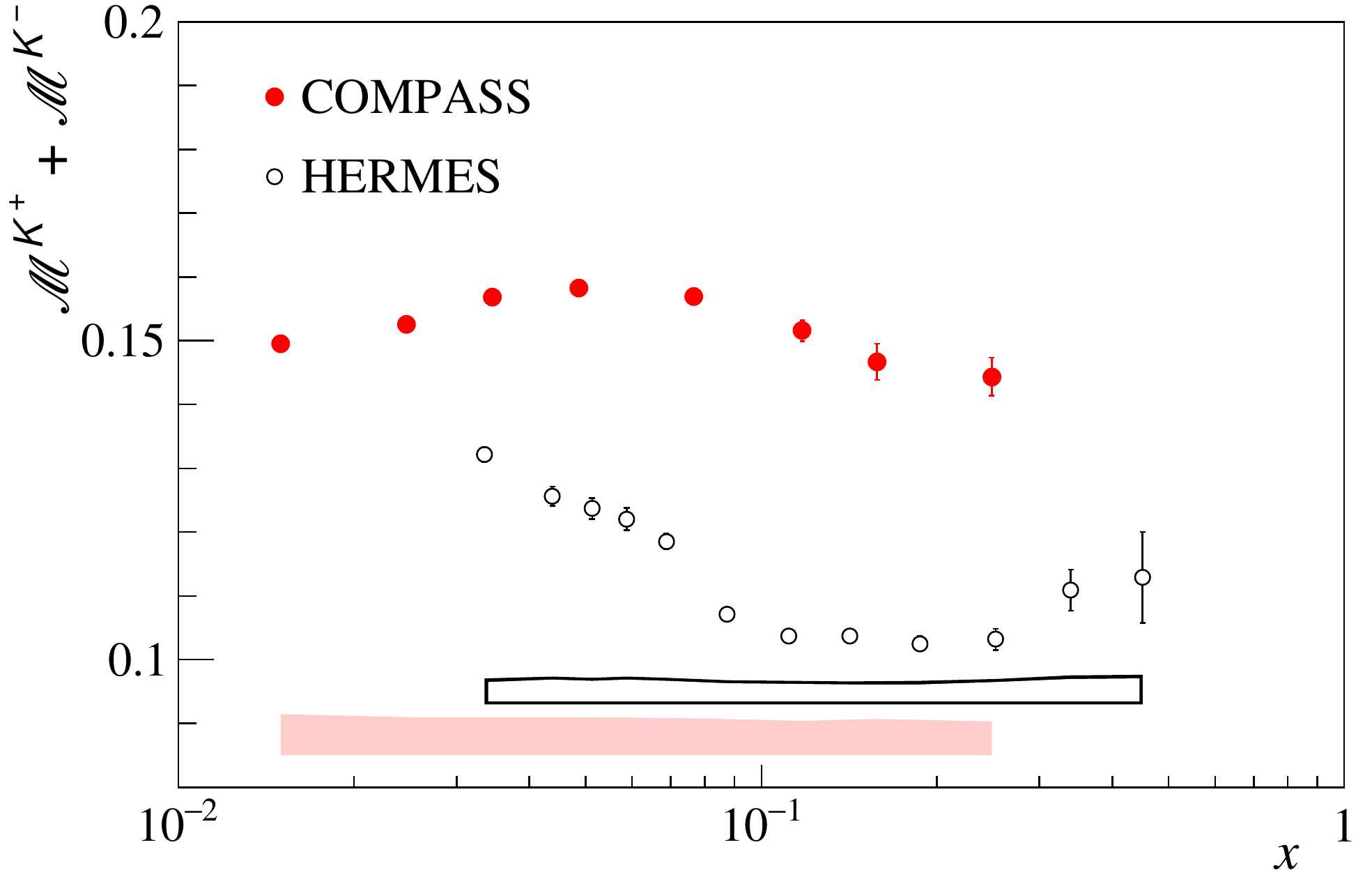}
\caption{Sum of $z$-integrated multiplicities, ${\mathscr M}^{\rm K^{+}}+{\mathscr M}^{\rm K^{-}}$. COMPASS data (full points) are compared to HERMES data~\cite{HERMESkaons} (open points). The bands show the total systematic uncertainties.
 }
\label{fig:intk}
\end{figure}

Since an important goal of the measurement of kaon multiplicities is the determination of quark-to-kaon FFs, the sum of positive and negative charged-kaon multiplicities is of special interest. It was used by HERMES \cite{HERMESkaons} for a LO pQCD extraction of the product of the strange quark PDF and the strange quark-to-kaon FF. For the COMPASS kinematics, the applicability of the LO pQCD formalism to pion multiplicities was already demonstrated~\cite{pionpaper}.
For an isoscalar target and assuming isospin and charge symmetry, the sum of $\rm K ^+$ and $\rm K^-$ multiplicities can be written in LO in a very simple way~\cite{HERMES2008}
. 
For multiplicities integrated over $z$, ${\mathscr M}^{\rm K^{\pm}}= 
\int \langle M^{\rm K^{\pm}}(x,y,z) \rangle_y\,{\text d}z$, the equation reads:

\begin{equation}
{\mathscr M}^{\rm K^+}+{\mathscr M}^{\rm K^-}
= \frac{U {\mathscr D}_U^{\rm K} + S {\mathscr D}_S^{\rm K}}{5U+2S},
\end{equation}
with $U=u+\bar{u}+d+\bar{d}$, and $S=s+\bar{s}$.
The $z$-integrated FFs, ${\mathscr D}^{\rm K}(Q^2)=\int D^{\rm K}(z,Q^2)\,{\text d}z $, depend on $Q^2$ only. The symbols ${\mathscr D}^{\rm K}_U$ and ${\mathscr D}_S^{\rm K}$ denote the 
combinations
of FFs 
\ ${\mathscr D}_U^{\rm K} = 4{\mathscr D}_{\rm u}^{\rm K^{+}}+4{\mathscr D}_{\rm \bar{u}}^{\rm K^{+}}+{\mathscr D}_{\rm d}^{\rm K^{+}}+{\mathscr D}_{\rm \bar{d}}^{\rm K^{+}}$ and ${\mathscr D}_S^{\rm K} = 2{\mathscr D}_{{\rm s}}^{\rm K^+}+2{\mathscr D}_{\rm \bar{s}}^{\rm K^+}  $.
At high values of $x$, the strange content of the nucleon can be neglected, and
in a good approximation the sum of K$^+$ and K$^-$ multiplicities is related to 
${\mathscr D}_U^{\rm K}/5$. 
This value is expected to have a rather weak $Q^2$ dependence
when integrated over $z$, so that it 
can be used at smaller values of $x$ to determine the $S {\mathscr D}_S^{\rm K}$ value.
%
The result for ${\mathscr M}^{\rm K^{+}}+{\mathscr M}^{\rm K^{-}}$ is presented in Fig.~\ref{fig:intk} as a function of $x$ at the measured values of $Q^{2}$.
The data are integrated over $z$ in the range 0.20 to 0.85 and averaged over $y$ in the range 
0.1 to 0.7. Only those eight $x$ bins which have a sufficient $z$ coverage are shown. A weak $x$ dependence is observed. Figure~\ref{fig:intk} also shows the HERMES results~\cite{HERMESkaons} that were taken at 27.6 GeV beam energy. 
They lie well below the COMPASS points and exhibit a different $x$ behaviour. 

From the COMPASS result on ${\mathscr M}^{\rm K^{+}}+{\mathscr M}^{\rm K^{-}}$ at high $x$ ($x=0.25$) we extract ${\mathscr D}_U^{\rm K} \approx 0.65-0.70$. This differs from the earlier DSS fit result at $Q^2=3\,(\GeV/c)^2$, ${\mathscr D}_U^{\rm K}= 0.43 \pm 0.04$~\cite{DSSFFs}, which was mainly based on HERMES results.
Towards low $x$, COMPASS data show a flat behaviour, unlike the rise that is suggested by the HERMES data and the DSS FF parametrisation\cite{DSSFFs}.
%

%

\begin{figure}[htbp]
\centering
\includegraphics[width=8.6cm]{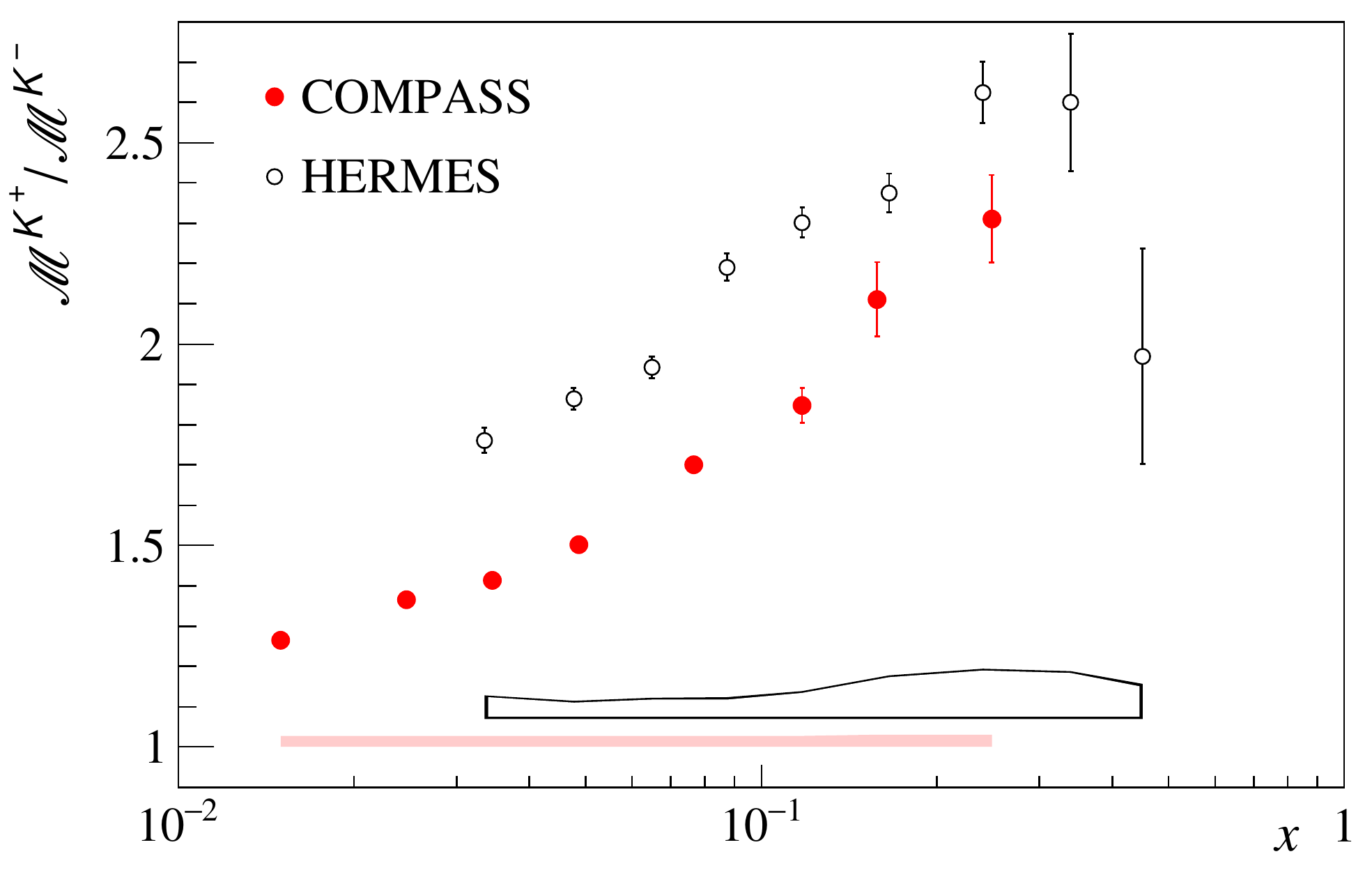}
\caption{Ratio of $z$-integrated multiplicities, ${\mathscr M}^{\rm K^{+}}/{\mathscr M}^{\rm K^{-}}$. COMPASS data (full points) are compared to HERMES data~\cite{HERMES2013} (open points). The bands show the total systematic uncertainties. 
}
\label{fig:ratio}
\end{figure}

Another quantity of interest is the $x$ dependence of the multiplicity ratio 
${\mathscr M}^{\rm K^{+}}/{\mathscr M}^{\rm K^{-}}$, in which most experimental 
systematic effects cancel. The results are shown in Fig.~\ref{fig:ratio} as a function of $x$. In the region of overlap, COMPASS results are found to be systematically lower than those of HERMES. In contrast, for the case of pions COMPASS and HERMES multiplicity ratios are found in good agreement~\cite{pionpaper}. 

\section{Summary and conclusions}
Precise results for charged-kaon multiplicities are obtained from kaon SIDIS measurements using a 
muon beam scattering off an isoscalar target. The data are given in a three-dimensional
$x$, $y$, and $z$ representation and cover a wide kinematic range : $1 \,(\GeV/c)^2< Q^{2} < 60\,(\GeV/c)^2$, $10^{-3} < x < 0.7$, $0.1 < y < 0.7$, and $0.20 < z < 0.85$ with $W > 5 $ GeV/$c^2$. They constitute an important input for future world-data analyses in order to constrain strange quark PDFs and FFs.
The sum of $\rm K^+$ and $\rm K^-$ multiplicities integrated over $z$ shows a 
flat distribution in $x$, with values significantly higher than those measured
by HERMES at lower energy. By covering lower $x$ values, the 
{multiplicity sum data} will significantly improve the constraint on the product of the 
strange quark PDF and FF, $SD_{\rm str}^{\rm K}$. 
{The present result points to a larger value of ${\mathscr D}_U^{\rm K}$ than obtained from the earlier DSS fit.}

\section*{Acknowledgements}
We gratefully acknowledge the support of the CERN management and staff and the
skill and effort of the technicians of our collaborating institutes.  This work
was made possible by the financial support of our funding agencies.  

\bibliographystyle{prsty}

\end{document}